\documentclass{article}

\usepackage{arxiv}

\usepackage[utf8]{inputenc} 
\usepackage[T1]{fontenc}    
\usepackage{url}            
\usepackage{booktabs}       
\usepackage{amsfonts}       
\usepackage{nicefrac}       
\usepackage{microtype}      
\usepackage{lipsum}
\usepackage{graphicx}
\usepackage{threeparttable} 
\usepackage{lineno,hyperref,graphicx,subfigure,url,amsmath}
\usepackage[section]{placeins}
\graphicspath{ {./images/} }

\title{Coronavirus statistics causes emotional bias: a social media text mining perspective}

\author{
 Linjiang Guo \\
  Institutes of Science and Development \\
  Chinese Academy of Sciences\\
  Beijing,100081 \\
  \texttt{scpaulgul@163.com} \\
   \And
 Zijian Feng \\
  School of Systems Science\\
  Beijing Normal University\\
    Beijing, 100875 \\
  \texttt{fxdwsfzj@126.com} \\
  \And
 Yuxue Chi \\
  School of Management Science and Engineering\\
  Central University of Finance and Economics\\
    Beijing, 100081 \\
  \texttt{chiy@cufe.edu.cn} \\
  \And
 Mingzhu Wang \\
  Institutes of Science and Development \\
  Chinese Academy of Sciences\\
  Beijing,100081 \\
  \texttt{wangmingzhu19@mails.ucas.ac.cn} \\
\And
 Yijun Liu*\\
  Institutes of Science and Development \\
  Chinese Academy of Sciences\\
  Beijing,100081 \\
  \texttt{yijunliu@casipm.ac.cn} \\
  \And
}

\begin{document}
\maketitle
\begin{abstract}
While COVID-19 has impacted humans for a long time, people search the web for pandemic-related information, causing anxiety. From a theoretic perspective, previous studies have confirmed that the number of COVID-19 cases can cause negative emotions, but how statistics of different dimensions, such as the number of imported cases, the number of local cases, and the number of government-designated lockdown zones, stimulate people's emotions requires detailed understanding. In order to obtain the views of people on COVID-19, this paper first proposes a deep learning model which classifies texts related to the pandemic from text data with place labels. Next, it conducts a sentiment analysis based on multi-task learning. Finally, it carries out a fixed-effect panel regression with outputs of the sentiment analysis. The performance of the algorithm shows a promising result. The empirical study demonstrates while the number of local cases is positively associated with risk perception, the number of imported cases is negatively associated with confidence levels, which explains why citizens tend to ascribe the protracted pandemic to foreign factors. Besides, this study finds that previous pandemic hits cities recover slowly from the suffering, while local governments' spending on healthcare can improve the situation. Our study illustrates the reasons for risk perception and confidence based on different sources of statistical information due to cognitive bias. It complements the knowledge related to epidemic information. It also contributes to a framework that combines sentiment analysis using advanced deep learning technology with the empirical regression method.
\end{abstract}

\section{Introduction}
    Undoubtedly, COVID-19 has become one of severe pandemics all over the world. It has caused hundreds of thousands of infected cases and Thousands of deaths daily\cite{WHO_2021}. The virus cannot be eliminated in the short term, therefore people will be forced to coexist with the Coronavirus in the post-pandemic era. Under such conditions, small-scale local epidemics caused by COVID-19 often attract great public attention\cite{zhong2021mental}. Former COVID-19 studies which concerns the role of information torch on issues including daily sentiment change monitoring algorithms\cite{lyu2020sentiment, barkur2020sentiment}, mental health toll\cite{li2020impact, zhong2021mental}, infodemic during pandemic\cite{pian2021causes}, and impact on education\cite{hofer2021online}. Scholars also explored the sources of information for panic about COVID-19 through questionnaires\cite{ahmad2020impact}. citizens' negative emotions put sever pressure on governments, leading to health policies changing drastically in America\cite{jin2021mining}.
    
    In this study, we are interested in the impact of coronavirus statistics on the emotional status, that is, individuals' risk perception of infection and confidence in eradicating the pandemic. A wealth of evidence demonstrates that bad emotions such as anxiety are attributed to an attentional bias as well as an information processing bias that operates to identify threats and risks when facing massive information \cite{mathews1994cognitive} \cite{hayes2007information}. However, the proportions of texts that express anxiety or confidence change differently with the pandemic severity, leading to a sneaking suspicion that various kinds of information play diverse roles in information processing and understanding \cite{lyu2020sentiment}. Understanding roles of sources of information is critical for advancing cognition research and improving the level of policy-making. 
    
    Meanwhile, Sentiment analysis, which is also called opinion mining, plays a part in understanding our society, especially in conjunction with social media texts. It aims to discover people's opinions on specific things such as a product, a company or government policy\cite{das2001yahoo,ouyang2022patients}. With the widespread use of the internet, everyone is now not only a listener to information but also a communicator of information as they can express their opinion in cyberspace freely. The consequences of text sentiment analysis may be more objective if there is sufficient data. Traditional methods of sentiment analysis on social media text are either dictionary- and rule-based methods \cite{rosenthal2017semeval, xiong2020understanding, guo2021modeling}, or shallow models that focus on carefully designed effective features\cite{chaturvedi2018bayesian}, which are tailored to obtain a satisfactory classification of opinion polarities. As for sentiment analysis technology, deep learning methods function better than traditional methods such as the emotional dictionary. Deep neural networks have many different structures and techniques, such as convolutional neural networks (CNNs), recurrent neural networks (RNNs), and attention mechanisms\cite{yadav2020sentiment}.

    Although existing studies using sentiment analysis have unveiled some interesting phenomena and put forward some valuable arguments, some potential problems still need to be studied. Firstly, some research has demonstrated that information has a significant impact on citizens' emotions during the pandemic\cite{ahmad2020impact} \cite{wang2022global}. But the understanding of why cyberspace turns to be in an uproar is not clear. Former research shows the proportions of positive emotion and negative emotion do not illustrate a significant trend changing with the number of cases, which indicates a complex impact mechanism\cite{lyu2020sentiment}. For instance, there are always intense discussions and polarization about whether the pandemic will end or not among China's online community. Discovering which kind of information leads to risk perception or confidence will be helpful to have a grasp of it. Next, from methodological perspectives, some research considers texts that are collected with hashtags such as "Coronavirus". These will ignore texts related to other aspects of the disease, such as masks \cite{kalampokis2013understanding}. Alternatively, if all social media texts are examined, there will be too much redundancy and noise\cite{lyu2020sentiment}. Additionally, there are many abbreviations and acronyms in Chinese social media texts\cite{xu2020deep}. Previous studies use emotional dictionaries or questionnaires to classify emotions, resulting in a potential bias in results\cite{ahmad2020impact}.

    Research demonstrates that perceived risk is related positively to affective responses to risk\cite{park2021optimistic}. Attentional bias theory states that people usually focus on different information alternatively when they seek emotional needs\cite{mathews1994cognitive}, on account of personal information processing bias \cite{hayes2007information}. A study presents two main factors that cause people to panic: fake news and the number of Coronavirus cases\cite{ahmad2020impact}. Therefore, we propose the following hypotheses based on the literature mentioned above.
    \begin{itemize}
    \item H1.a The number of local cases is positively related to risk perception.
    \item H1.b The number of local cases does not have an effect on confidence levels.
    \end{itemize}
     
     Regarding policy, in some countries and regions, such as mainland China, Hong Kong and Taiwan, the CDC divides cases into two parts: local cases and imported cases, which may cause an attentional bias when residents seek health information. As far as our information goes, imported cases will be quarantined in a fixed location, but they may lead to an endless pandemic\cite{russell2021effect}. Changes in emotions shown in the supplementary material do not seem related. Hence, we set out the following hypotheses: 
    
    \begin{itemize}
    \item H2.a The number of imported cases are negatively related to confidence levels.
    \item H2.b The number of imported cases does not have an effect on negatively related to risk perception.
    \end{itemize}

    Based on bias cognition theory, this study develops hypotheses and test whether people pay attention to different sources of information when searching for it to estimate personal risk or future state\cite{mathews1994cognitive}. In other words, coronavirus statistics impacts fall into two main categories. Moreover, this study proposes an infusion deep neural network named attention-based channels-LSTM multitask-learning model (ACLMM) to carry out a multi-class sentiment analysis, leading to outcomes suited for regression analysis. In the empirical part, we employ a fixed-effects estimation after model tests, for which the results are rigorous and credible.

\section{Background}
    \subsection{Social malaise caused by pandemics}
    
    There have been many applications of sentiment analysis and affective computing in regard to pandemics. An early study has shown that people who feel panic are more likely to behave irrationally during a pandemic\cite{fast2015modelling}. Irrational behaviors include expressing stress, escaping from the outbreak site, and challenging official announcements. These behaviors may even lead to recession\cite{cheng2004paranoid}. A previous study confirmed that outbreaks of disease hurt the mood of internet users\cite{valdez2020social}. \cite{lwin2020global} examined worldwide trends of different emotions during the COVID-19 pandemic. \cite{li2020impact} found that negative emotions increased, while positive emotions and life satisfaction declined during the first wave of the Coronavirus pandemic. \cite{basiri2021novel} discovered that Coronavirus attracted the attention of a wide range of people at different times in varying intensities. A study determined two main factors that cause people to panic: fake news and the number of Coronavirus cases\cite{ahmad2020impact}. However, surprising risk information may change beliefs and improve the accuracy of risk perception\cite{sinclair2021pairing}. As the pandemic moved into different countries, it affected the countries' financial markets markedly\cite{ali2020coronavirus}. 
   
    Negative emotions cause much harm during pandemics, and researchers can use sentiment analysis to reveal public concerns and panic during the pandemic. The emotions are possible mediating variables in the relationship between COVID-19 pandemic and different fields such as tourism\cite{karl2021affective}, politics\cite{druckman2021affective}, education\cite{roman2021effectiveness}, vaccination\cite{doi:10.1073/pnas.2118721119}, and finance\cite{barrafrem2021trust}. To handle hardships, some research stress the importance of governments. High-trust societies which could perform well in pandemic rely on a reliable and professional bureaucracy, a good economic situation, and low population density\cite{christensen2020balancing}.
    
    In terms of the post-pandemic era, there is so much uncertainty and complexity. Periods after the first pandemic wave are called post-pandemic. A study finds that the proportion of infections belonging to the elderly is particularly small during periods of low prevalence and continuously increases as case numbers increase\cite{steinegger2021behavioural}. The authors model the dynamical adoption of prophylaxis through a two-strategy game and couple it with a SIR spreading model. Some studies discuss the change of world economy and the new challenges of globalization in the post-pandemic era. For instance, \cite{kolodko2020after}'s article proposes that the changes induced by heightened nationalism and protectionism will be marginal rather than fundamental. In contrast, the future world economy will need even more globalization. Transformations are needed in science-policy, economies and governance between post-pandemic\cite{leach2021post}. Mental loss has existed for long time after a disaster. Research finds that the level of morbidity of posttraumatic stress disorder four months after a fire disaster remained almost unchanged at 29 months, indicating the long-term nature of posttraumatic stress\cite{mcfarlane1988aetiology}. College students who self-reported COVID-19 infections were more likely to experience anxiety and depression\cite{goldrick2022self}.
    
    \subsection{Sentiment analysis}
   
    Traditional sentiment analysis is a binary classification task, while affective computing focuses on extracting a set of emotion labels\cite{cambria2017affective}. A direct and traditional technique for text sentiment analysis uses a sentiment analysis dictionary, which calculates the sentiment polarity score with a specific algorithm, such as the naïve Bayes method, after counting the frequency of sentimental words\cite{kang2012senti}. Conjunction words and adjectives with comparative forms can provide the polarity of emotion, researchers usually choose adjectives as keywords of emotional dictionaries \cite{hatzivassiloglou1997predicting} \cite{hatzivassiloglou2000effects}. It is also difficult to remove emotional ambiguity according to the context. Therefore, to address the drawbacks mentioned above, some studies use a word vector to construct the dictionary in order to obtain a better word representation\cite{zhang2020does}.

    Inspired by biological neurons, an artificial neural network abstracts neuronal computations into a simple mathematical form briefly referred as the "M-P" neuron\cite{mcculloch1943logical}. A neural network with only one neuron can be regarded as a double-layer neural network with only an input layer and an output layer and no hidden layer.The structure of a neural network can be divided into three parts: an input layer, an output layer and the hidden layers. The output of each neuron can be written as $f(x)=W_ix_i+b$. A perceptron algorithm describes the way neurons process values. Based on duality theory, this algorithm proposes an iterative method to calculate the weight W and offset b, namely, stochastic gradient descent (SGD). Therefore, the perceptron algorithm can be regarded as a special two-layer neural network. For a long time previously, due to the limitation of computing power, a network with a few layers was considered practical. However, the situation has changed greatly with increases in computing power, which has enabled deep neural networks with many layers and complex architectures to achieve amazing results.

    Deep learning methods often perform better than traditional methods in sentiment analysis. To address the drawback of the absence of contextual information, learning word embedding is a way to obtain text features, and word embedding rather than it-idf has been used in many recent works\cite{hinton1986learning},\cite{li2018word}. Convolution neural networks (CNNs) belonging to deep learning methods are a kind of special feedforward neural network that were originally used in the field of computer vision. Variants of CNNs used for sentiment analysis applications, such as AC-BiLSTM, handle differences between sentences through multiple LSTM layers\cite{liu2019bidirectional}. LSTM networks are a type of recurrent neural network (RNN) capable of learning order dependence in sequence prediction problems\cite{hertz2018introduction}. LSTM can be used as an encoder to embed words that act as a one-hot vector, so that the word vector can learn the dependency relationship over a longer distance\cite{johnson2016supervised}. To solve the problem of long-distance information loss, a multilayer LSTM or bidirectional LSTM can be constructed, which updates only some of the parameters in each training epoch\cite{liu2015multi},\cite{jaseena2021decomposition}.

    Although the schemes mentioned above should be mostly adequate to handle long-distance dependencies in the text, the overall performance of their sentiment analysis is still insufficient. Researchers have proposed two methods for improvement: one is an attention mechanism to learn long-distance information, and the other is multitask learning that learns the language rules. The attention mechanism is inspired by the visual attention mechanism in humans. Humans pay high attention to specific things and reduce attention to other things at the same time. For instance, we can hear our partner's voice even in a noisy environment. 
    
\begin{figure}[tbh]
        \centering\includegraphics[scale=1.0]{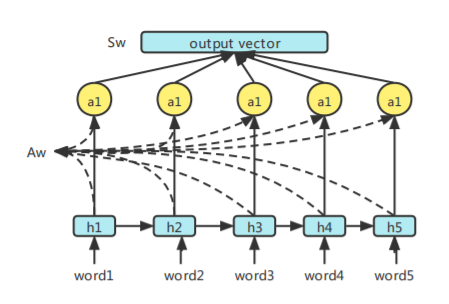}
        \caption{Attention mechanism}
        \label{fig:att}
    \end{figure}

    In a natural language processing task, the attention mechanism makes the neural network know what is important based on the input text (i.e. key-value matching) \cite{vaswani2017attention}. As \autoref{fig:att} illustrates, $A_{w}$ represents the weight of attention imposed on sum of weights multiplied by input, which leads to less loss when it is long-history information. The attention mechanism was first applied in machine translation tasks\cite{bahdanau2014neural}. The attention mechanism has also been widely used in the field of sentiment analysis \cite{zhou2016attentioncross},\cite{du2017convolutional},\cite{cai2020recurrent}. In addition, the attention mechanism can also be applied between labels and text matching\cite{wang2018joint}, as well as in aspect-based sentiment analysis\cite{trueman2021convolutional}.
    
    Multitask Learning (MTL) is a learning paradigm in machine learning methods. Its purpose is to exploit the useful information contained in multiple related tasks to help improve the performance of all tasks\cite{zhang2021survey}. MLT can be combined with other learning techniques, including semisupervised learning, supervised learning, unsupervised learning and reinforcement learning. MTL can be used to further improve the performance of a model\cite{liu2016recurrent}. Besides, many new models such as Transformer and graph-based representations, have been used in recent years\cite{linmei2019heterogeneous},\cite{tang2020dependency}.

\section{Material and methods}    
\begin{figure}[tbh]
        \centering\includegraphics[scale=0.5]{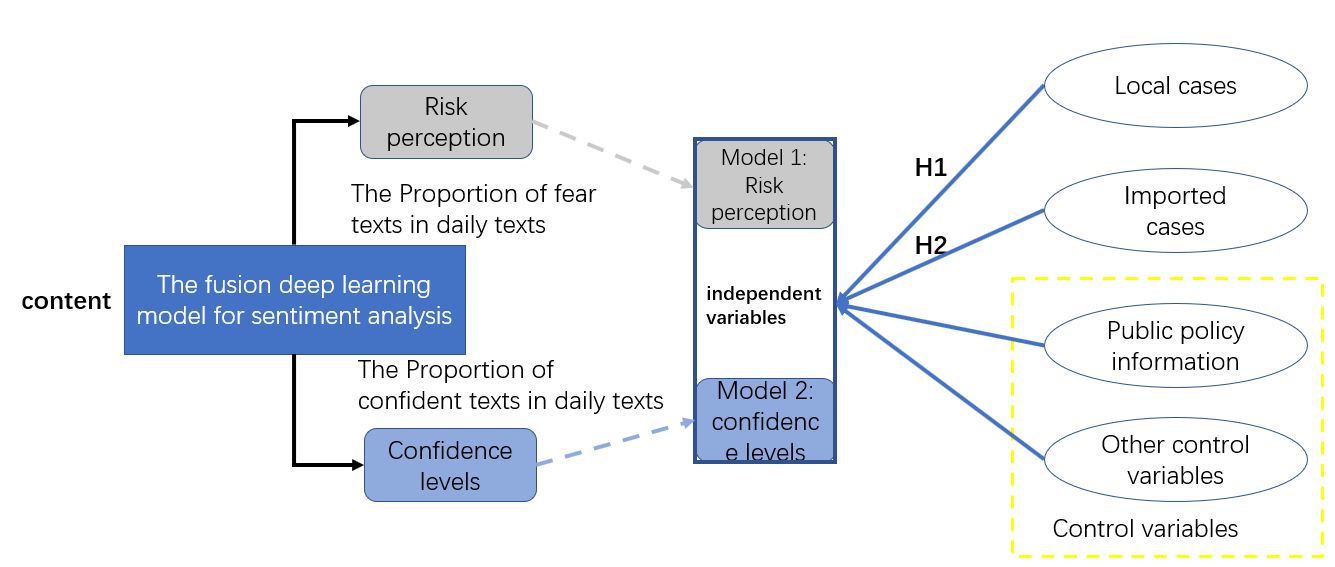}
        \caption{Illustration of the Framework for the Study}
        \label{fig:framework}
    \end{figure}

    In this study, we are interested in how different coronavirus statistics affect public opinion, specifically, risk perception and confidence levels. But a massive amount of unstructured text data hamper our further investigation. To handle the problem, we first propose a deep multi-class sentiment analysis model to find and analyze texts related to the pandemic. Then we conduct a panel data analysis with a fixed-effect model. The study establishes a connection between deep learning methods and econometric methods, the framework of it is illustrated in \autoref{fig:framework}.

\subsection{Proposed multi-class sentiment analysis model} 
    In this section, we describe the detailed architecture of ACLMM. CNNs’ characteristics are using kernels to scan the text vector-matrix so that the neural network can learn local information. The local information includes metadata, slang phrases, and emoticons, which is well suited to Chinese social media texts\cite{xu2020deep}. On the other hand, RNNs, such as long short-term memory (LSTM) and gated recurrent unit networks, can learn global information\cite{cho2014learning}. Although the networks are capable of solving the problem of long-distance dependency to some extent, their learning ability will be impaired if the sentences are too long. So we use the attention mechanism to attach importance to keywords. Moreover, the learning process is based on multi-task learning because a single Chinese corpus is small.

    The architecture of ACLMM is as following: firstly, converting the text to word vector matrix as input of convolutional layer. Suppose that the shape of the word vector matrix is $W_{s,d}$, where $s$ is the length of text and $d$ is the length of each word vector. The size of the kernel is $k$, the number of channels is $c$. Concatenating different outputs of channels that are in the same position and gaining $S_{s-d+1,c}$. And then the outputs are used as the input of a bi-directional LSTM network in which the length of the hidden unit is $c$. In a specific task, copying the hidden unit of bi-directional LSTM layer as input of a specific LSTM layer that length of the hidden unit is $2c$. The classification result is obtained by attention mechanism and softmax operation on the output after the final unit of the specific LSTM layer. The ACLMM is as follows \autoref{fig:ACLMM}. We give detail information of the model in the supplementary material section 1.

    \begin{figure}[tbh]
        \centering\includegraphics[scale=0.7]{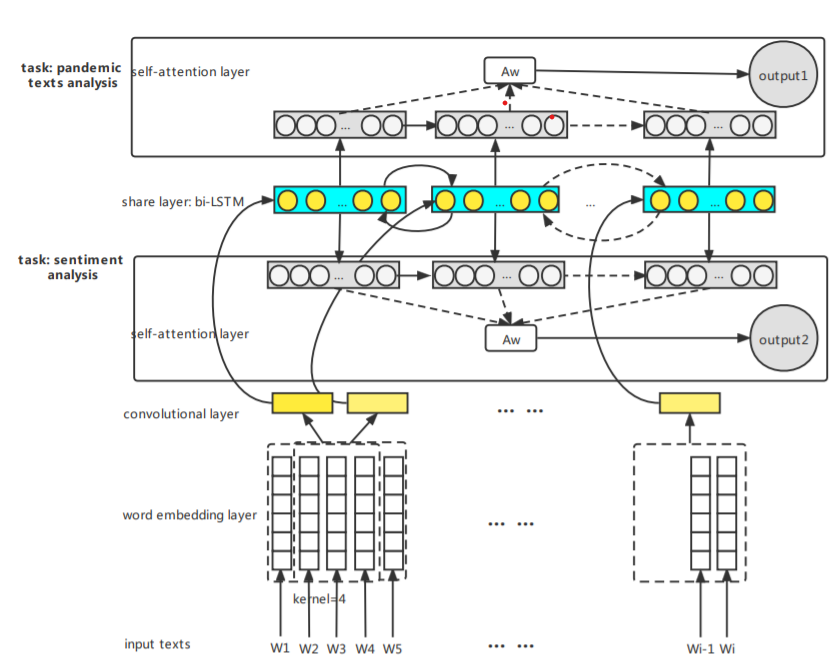}
        \caption{Architecture of ACLMM}
        \label{fig:ACLMM}
    \end{figure}
    
    The attention mechanism calculates the output of parallel tasks' LSTM layer. The calculation formula is as follow:
    $${\alpha}_{t}=\frac{\exp (\mathbf{v} \cdot \mathbf{h_{t}})}{\sum_{i = 1}^{\infty} \exp (\mathbf{v} \cdot \mathbf{h_{i}})}$$
    $$\mathbf{S}_{\mathbf{A}_{\mathbf{w}}}=\sum_{t} {\alpha}_{t} \mathbf{h}_{t}$$
    Where $v$ is the output of the final timestep of LSTM and ${\alpha}_{t}$ is the weight of each word. The category of output is based on the softmax function. The details of the model are as following:

\subsection{algorithm details}
 This study uses the CBOW algorithm to train word vectors based on a large-scale corpus that sourced from Natural Language Processing and Information Retrieval sharing platform\footnote{http://www.nlpir.org/wordpress/}. We choose the CBOW algorithm because it has a faster training speed. Former research demonstrates that character-level word vector performances better than word-level word vector in several tasks based on Chinese social media\cite{xu2020deep}. So, we deal with the original corpus according to the character segmentation and train word vectors. Finally, we obtain a word vector matrix with a width of 200-dimension and a length of 6,644(words).
 
    The layer of convolution with multi-channel is used to detect emotion from different aspects. The vectors extracted from the same position but different channels are concatenated to produce a feature map$M=\left[\overrightarrow{m_{0}} , \overrightarrow{m_{1}} , \ldots, \overrightarrow{m_{s-k+1}}\right]$, where s is the length of the text and k is the size of the kernel. The length of $\overrightarrow{m_{i}}$ is equal to the number of channels.

    Bi-directional LSTM consists of cells with cell state and gates mechanism\cite{graves2013generating}. Cell state$C_{t}$ depends on the former time cell state$C_{t-1}$ and hidden state$h_{t-1}$, current time input$X_{t}$. Its calculation formula is as follows:
    $$\mathbf{c}_{t}=\mathbf{f}_{t} \odot  \mathbf{c}_{t-1}+\mathbf{i}_{t} \odot \tanh \left(\mathbf{W}_{x c} \mathbf{x}_{t}+\mathbf{W}_{h c} \mathbf{h}_{t-1}+\mathbf{b}_{c}\right)$$
     LSTM adopts the gating mechanism and has three types of gates. All gates are parametric functions that can be trained. First is the input gate, it deals with the former time information and current time input information. The calculation formula is as follows: $$\mathbf{i}_{t}=\sigma\left(\mathbf{W}_{x i} \mathbf{x}_{t}+\mathbf{W}_{h i} \mathbf{h}_{t-1}+\mathbf{W}_{c i} \mathbf{c}_{t-1}+\mathbf{b}_{i}\right)$$ 
    Forget gate decides how the former time cell influences the current cell's memory by controlling forget some information. The calculation formula is: $$\mathbf{f}_{t}=\sigma\left(\mathbf{W}_{x f} \mathbf{x}_{t}+\mathbf{W}_{h f} \mathbf{h}_{t-1}+\mathbf{W}_{c f} \mathbf{c}_{t-1}+\mathbf{b}_{f}\right)$$
    Output gate transmits information to the next cell and decides which parts of the cell state should be outputted according to the following equations.
    $$\mathbf{o}_{t}=\sigma\left(\mathbf{W}_{x o} \mathbf{x}_{t}+\mathbf{W}_{h o} \mathbf{h}_{t-1}+\mathbf{W}_{c o} \mathbf{c}_{t}+\mathbf{b}_{o}\right)$$
    $$\mathbf{h}_{t}=\mathbf{o}_{t} \odot \tanh \left(\mathbf{c}_{t}\right)$$
    bi-directional LSTM can obtain more information by learning emotional information from two opposite directions.

    Each task has its layers: an LSTM and an attention layer. The hidden states of the bi-directional LSTM layer are copied to each task's LSTM layer. As for the bi-directional LSTM layer, the output of a cell from front to back is $\overrightarrow{h_{front,i}}$, it is $\overrightarrow{h_{back,i}}$ from back to front. So the input of a cell in the each task's uni-directional LSTM is $\overrightarrow{h_{front,i}} \oplus \overrightarrow{h_{back,i}}$. The attention mechanism assigns different weights to words according to different contributions to the emotion. A common way to assign weights to calculate the similarity between hidden state and final output with the vector product is the attention mechanism. In this model, the attention mechanism calculates the output of parallel tasks' LSTM layer. The calculation formula is as follow:
    $${\alpha}_{t}=\frac{\exp (\mathbf{v} \cdot \mathbf{h_{t}})}{\sum_{i = 1}^{\infty} \exp (\mathbf{v} \cdot \mathbf{h_{i}})}$$
    $$\mathbf{S}_{\mathbf{A}_{\mathbf{w}}}=\sum_{t} {\alpha}_{t} \mathbf{h}_{t}$$
    Where $v$ is the output of the final timestep of LSTM and ${\alpha}_{t}$ is the weight of each word. The category of output is based on the softmax function.

\subsection{Training datasets}

    Six Chinese datasets about sentiment analysis are used in experiments, and the description of datasets is shown in the \autoref{tab:dataset}. The first five datasets are derived from previous research works or open-source platforms. This work labels the last dataset.
    \begin{table}[]
        \centering
        \caption{description of datasets}
        \resizebox{\columnwidth}{!}{%
        \centering\begin{tabular}{lp{8cm}p{2cm}p{2cm}}
        \hline
        Dataset  & Description       & Number of categories \\ \hline
        Nnline\_shopping\_10\_cats\cite{zhang2020long} & More than 60,000 comments data about ten categories of commodities, including positive and negative comments about 30,000    & 2    \\
        Simplifyweibo\_4\_moods    & There are about 360,000 emotional messages on Sina Weibo, of which 200,000 are joy, and 50,000 are anger, disgust and depression. & 4    \\
        weibo\_senti\_100k         & More than 100,000 emotional texts on Sina Weibo, with about 50,000 positive and 50,000 negative comments                       & 2    \\
        Nlpcc2014\_sentiment\footnote{http://tcci.ccf.org.cn/conference/2014/}       & More than 10,000 emotional texts, divided into three types: positive, negative and neutralize                      & 3    \\
        COVID-19\_sentiment\cite{lyu2020sentiment}        & There are over 20,000 social media texts with seven emotional attributes about COVID-19 after we integrate some neutral and ambivalent texts as an uncertain type. Categories are fear, disgust, joy, surprise, confidence, sadness, anger, and uncertainty               & 7    \\
        COVID-19\_identify         & This study collected and labeled around 9,000 texts from Sina Weibo, half of them were related to COVID-19, and half of them were not.    & 2     \\ \hline
        \end{tabular}
        }
        \label{tab:dataset}
    \end{table}

    In this part, we use there criteria namely accuracy($A_{cc}$), precision($P_{r}$), and recall($R_{e}$) to assese the performance of the algorithm. Accuracy is used to measure whole classification performance. we use both of them to measure performance in each single class because Precision and recall are contradictory in some cases. These criteria are extensively used in text classification and sentiment analysis tasks
    They are calculated as follows:

    $$A c c=\frac{T P+T N}{T P+F P+T N+F N}  $$
    $$P r=\frac{T P}{T P+F P}$$
    $$R e=\frac{T P}{T P+F N}$$
    $$F 1=\frac{2 \times P r \times R e}{P r+R e}$$
    where $T P$, $T N$, $F P$, and $F N$ are true positive, true negative, false positive, and false negative, respectively.

    \subsection{Model outputs} 
    The supplementary material(section 1) goes into detail about training datasets, training process, and evaluation. The result based on annotated COVID-19\_sentiment dataset compared with previous studies shown in \autoref{tab:score_table} and the training result shown in \autoref{fig:Acc and loss}, the ACLMM achieved better performance in the dataset. Compared with AC-BiLSTM\cite{liu2019bidirectional} that has multiple LSTM layers and attention layers, We find it is substantial to use the kernel structure in the task of Chinese informal text sentiment classification, which is in consistence with former research\cite{xu2020deep}. The MTL has significantly improved the performance of the task. Here we gain some outputs in the following formulas: 
 
\begin{figure}[tbh]
        \centering
        \begin{minipage}[t]{0.48\textwidth}
        \centering
        \includegraphics[width=6cm]{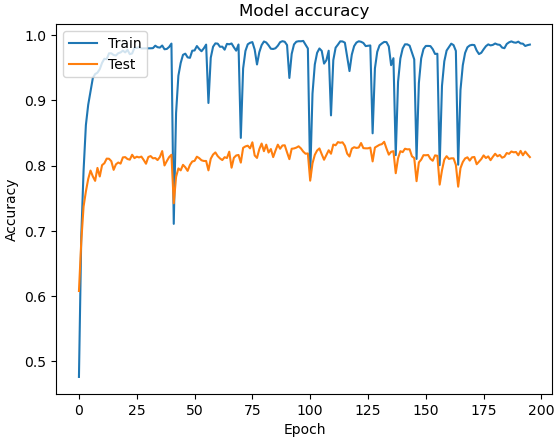}
        \end{minipage}
        \begin{minipage}[t]{0.48\textwidth}
        \centering
        \includegraphics[width=6cm]{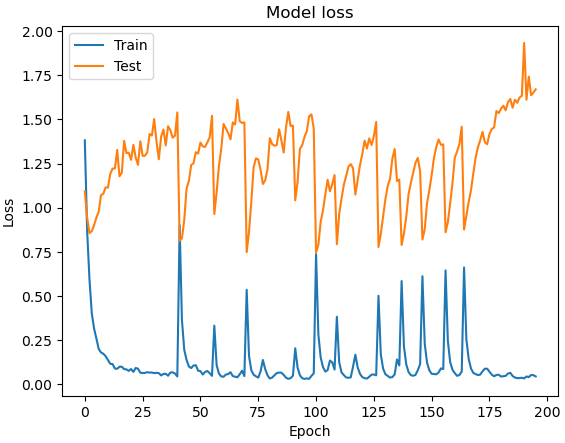}
        \end{minipage}
        \caption{Accuracy and loss changes in training}
        \label{fig:Acc and loss}
\end{figure}
    
    $$
    attention_{it} = \frac{A_{it}}{W_{it}}
    $$
    $$
    emotion_{it} = \frac{E_{it}}{A_{it}}
    $$
    where $A_{it}$, $W_{it}$ means the total number of texts and the number of texts belonging to the type of attention in day $t$ from city $i$, respectively. $emotion_{it}$ represents the number of a specific(e.g. fear) texts in day $t$ from city $i$.
    We call $attention_{it}$ as \textit{pandemic texts}, whose accuracy of the task that classifies pandemic texts is 86.02\%.
    Similarly, The precision and recall score of $emotion_{it}$ are shown in \autoref{tab:subdivide}. There are seven types of emotion, and it performs pretty well in the class of fear and confidence.
    
\begin{table}[tbh]   
        \centering
        \caption{Results in different models}
        \begin{tabular}{@{}lllllll@{}}
        \hline
            Name                                     & Accuracy     \\ \hline
            LSTM\cite{lyu2020sentiment}             & 0.78         \\
            Text-CNN\cite{zhang2015sensitivity}     & 0.7692       \\
            bi-LSTM                                  & 0.8011       \\
            AC-BiLSTM\cite{liu2019bidirectional}    & 0.8210       \\
            ACLMM\_without\_MTL                        & 0.8137       \\
            ACLMM                                    & 0.8418       \\ \hline
        \end{tabular}
        \label{tab:score_table}
    \end{table}
    
\begin{table}[]
    \centering
    \caption{Comparison of the results in each emotion}
    \begin{tabular}{llll}
    \hline
    Class      & Precision & Recall & F1     \\ \hline
    Fear       & 0.8987    & 0.9405 & 0.9191 \\
    Disgust    & 0.6895    & 0.6102 & 0.6475 \\
    joy       & 0.8427    & 0.7845 & 0.8126 \\
    surprise   & 0.8788    & 0.8985 & 0.8885 \\
    confidence & 0.9570    & 0.9727 & 0.9648 \\
    Sadness    & 0.7197    & 0.8123 & 0.7632 \\
    Anger      & 0.7947    & 0.7579 & 0.7759 \\ 
    Uncertainty& 0.8251    & 0.7992 & 0.8119 \\ 
    \hline
    \end{tabular}
    \label{tab:subdivide}
\end{table}

\subsection{Empirical methodology}

    Our analysis uses the data of 10 large cities in China over 31 days(Aug 1 - Aug 31), resulting in 462 observations. Since the data consists of a panel of 10 cities for 31 days, where N=10 is much less than T=31, the GMM estimator is not appropriate for our analysis. Definitions of variables are furnished in \autoref{tab:variables}. 
    There are three independent variables: \emph{attention}, \emph{fear}, and \emph{confidence}. All of the above variables are calculated according to their proportion in the studied texts based on the chosen categories. Attention is the proportion of COVID-19-related texts to the total number of texts in a city's social media communications in one day. Fear and confidence are the proportions of fear-related texts and confidence-related texts in a city's COVID-19-related texts on one day.
    
    \textbf{control variables:} We also try to analysis the historic impacts of each city, so we adopt some control variables to get rid of covariates. A limited government indicates that a government is limited in power. In contrast, A larger government has more government capacity and, correspondingly, more fiscal spending. We choose two indicators as the government's capacity: the number of urban health investments per capita and the amount of urban public service expenditures per capita. The spread of virus may be correlated with population density, which will also be a factor affecting people's emotions, so we choose urban population density as the control variable, which is determined from the wind database. The description of variables is provided in \autoref{tab:variables}

    First, we conduct a series of preliminary tests to choose a variety of regression model that is well-suited to our study\cite{asteriou2021public}. An F-test is conducted to determine whether a pool model should be used. The result rejects the null hypothesis of the pool model($F = 3.5477, p < 3.285e-09 $). considering exogeneity that includes the different properties of each city, we choose a pannel model is appropriate. Wald tests are employed to examine heteroskedasticity and autocorrelation in the data\cite{greene2000mmpi} \cite{wooldridge2002inverse}. The Wald test cannot reject the null hypothesis that there is no heteroskedasticity or autocorrelation in the data. Then the Breusch-Pagan LM test illustrates a simultaneous correlation between groups($ z = 9.3086, p-value < 2.2e-16 $)\cite{greene2000mmpi}. 
    
    Therefore, we use the robust covariance matrix vcovHC ( also known as the sandwich estimator) to employ the Hausman test because of heteroskedasticity and autocorrelation \cite{white1980heteroskedasticity}, \cite{white1984nonlinear}($ chisq = 11.098, p-value = 0.02548 $). Consequently, we use the fixed-effects panel model. An equation is as follows:

\begin{eqnarray*}
    {y_{c t}}=&\beta_0+\beta_1 \ln({cases_i}_{t-1})+\beta_2 \ln({foreign_i}_{t-1})+\beta_3 \ln({risk_i}_{t-1})+\beta_4 {distance_i}_{t-1}\\
            &+\beta_5 {pmedical_i}+\beta_6 {pgovernment_i}+\beta_7 {density_i}+\gamma_i D_i+\delta t_i+{\epsilon_i}_t 
\end{eqnarray*}


    Where $y_c t$ representing a category of c(i.g. fear or confidence) on day $t$. These variables are also described in \autoref{tab:variables}. In addition, $D_i$ is the between-group difference of city $i$, $t_i$ is a time trend term, ${\epsilon_i}_t$ is residual term.

\begin{table}[]
    \centering
    \caption{Explanation of the variables}
    \resizebox{\columnwidth}{!}{%
    \centering\begin{tabular}{lp{6cm}p{4cm}p{2cm}}
    \hline
    Variable name   &  Explanation      &  Data source & Unit          \\ \hline
    Attention & the proportion of texts in relation to pandemic & output of ACLMM & --- \\ 
    fear & the proportion of texts which show personal Concerns & output of ACLMM & --- \\ 
    confidence & the proportion of texts which show confidence on the pandemic & output of ACLMM & --- \\ 
    population outflow & net outflow of the population daily & Baidu Population Migration Data Platform & person \\ 
    cases   & The number of local new patients           & Government releases report &  person        \\ 
    Foreigners     & The number of patients from abroad          & Government releases report   & person        \\ 
    Risks   & The number of medium-risk areas and high-risk areas           & Government releases report   &     ---          \\ 
    Distance    & Distance between the current city and nearest one with patients    & Google Maps & km   \\ 
    Pmedical   & The per capita cost of health care for urban residents    & Chinese Yuan/per person   \\ 
    Pgovernment       & The per capita cost of public service expenditures    & Wind-Economic Database          & Chinese Yuan/per person   \\ 
    Density         & Urban population density           &  Wind-Economic Database    & population/sq.km. \\ \hline
    \end{tabular}
}
\label{tab:variables}
\end{table}     
    
\section{Results}
\subsection{Combining empirical analysis results}

\begin{table}[]
\caption{Regression analysis results}
\begin{threeparttable} 
\resizebox{\columnwidth}{!}{%
\begin{tabular}{lp{2cm}p{2cm}p{2cm}p{3cm}}
    \hline
    Variable Y &
      Model 1 (DV: Fear) &
      Model 2 (DV: Confidence)  &
      Model 3 (DV: Attention) &
      Model 4 (DV: Population outflow)(robust test)\\ \hline
    $Cases_{i, t-1}$ & 
      \begin{tabular}[c]{@{}l@{}}0.405***\\ (0.103)\end{tabular} &
      \begin{tabular}[c]{@{}l@{}}0.016\\ (0.094)\end{tabular} &
      \begin{tabular}[c]{@{}l@{}}0.788***\\ (0.130)\end{tabular} &
      \begin{tabular}[c]{@{}l@{}}0.013**\\ (0.006)\end{tabular} \\
    $Foreign_{i, t-1}$ &
      \begin{tabular}[c]{@{}l@{}}0.066\\ (0.124)\end{tabular} &
      \begin{tabular}[c]{@{}l@{}}-0.326***\\ (0.108)\end{tabular} &
      \begin{tabular}[c]{@{}l@{}}0.341***\\ (0.093)\end{tabular} &
      \begin{tabular}[c]{@{}l@{}}-0.004\\ (0.006)\end{tabular} \\
    $Risk_{i, t-1}$ &
      \begin{tabular}[c]{@{}l@{}}0.136**\\ (0.067)\end{tabular} &
      \begin{tabular}[c]{@{}l@{}}0.031\\ (0.055)\end{tabular} &
      \begin{tabular}[c]{@{}l@{}}0.412***\\ (0.103)\end{tabular} &
      \begin{tabular}[c]{@{}l@{}}-0.003\\ (0.002)\end{tabular} \\
    $Distance_{i, t-1}$ &
      \begin{tabular}[c]{@{}l@{}}0.013\\ (0.020)\end{tabular} &
      \begin{tabular}[c]{@{}l@{}}0.003\\ (0.019)\end{tabular} &
      \begin{tabular}[c]{@{}l@{}}0.016\\ (0.023)\end{tabular} &
      \begin{tabular}[c]{@{}l@{}}-0.001\\ (0.001)\end{tabular} \\
    $Pmedical_{i}$ &
      \begin{tabular}[c]{@{}l@{}}-0.008*\\ (0.004)\end{tabular} &
      \begin{tabular}[c]{@{}l@{}}0.013***\\ (0.004)\end{tabular} &
      \begin{tabular}[c]{@{}l@{}}0.005\\ (0.004)\end{tabular} &
      \begin{tabular}[c]{@{}l@{}}-0.002***\\ (0.000)\end{tabular} \\
    $Pgovernment_{i}$ &
      \begin{tabular}[c]{@{}l@{}}0.001\\ (0.004)\end{tabular} &
      \begin{tabular}[c]{@{}l@{}}-0.012**\\ (0.005)\end{tabular} &
      \begin{tabular}[c]{@{}l@{}}-0.002\\ (0.004)\end{tabular} &
      \begin{tabular}[c]{@{}l@{}}0.002***\\ (0.000)\end{tabular} \\
     $density_{i}$ &
      \begin{tabular}[c]{@{}l@{}}-2.022\\ (3.078)\end{tabular} &
      \begin{tabular}[c]{@{}l@{}}-7.028**\\ (3.220)\end{tabular} &
      \begin{tabular}[c]{@{}l@{}}-3.018\\ (2.507)\end{tabular} &
      \begin{tabular}[c]{@{}l@{}}0.272\\ (0.272)\end{tabular} \\
    \hline
\end{tabular}}
\begin{tablenotes}
    \footnotesize
    \item[1] Standard errors are in parenthesis
    \item[2] *** p\textless{}0.01, ** p\textless{}0.05, * p\textless{}0.1
\end{tablenotes}            

\end{threeparttable}       
\label{tab:combine_R}
\end{table}

\begin{table}[]
\caption{Intercept term}
\begin{threeparttable} 
\resizebox{\columnwidth}{!}{%
\begin{tabular}{lp{2cm}p{2cm}p{2cm}p{3cm}}
    \hline
     Variable Y &
      Model 1 (DV: Fear) &
      Model 2 (DV: Confidence)  &
      Model 3 (DV: Attention) &
      Model 4 (DV: Population outflow)(robust test)\\ \hline
 Beijing &
      \begin{tabular}[c]{@{}l@{}}-0.573\\ (3.456)\end{tabular} &
      \begin{tabular}[c]{@{}l@{}}11.210**\\ (4.358)\end{tabular} &
      \begin{tabular}[c]{@{}l@{}}1.610\\ (3.869)\end{tabular} &
      \begin{tabular}[c]{@{}l@{}}-1.812***\\ (0.374)\end{tabular} \\
    Chengdu &
      \begin{tabular}[c]{@{}l@{}}-6.860**\\ (2.775)\end{tabular} &
      \begin{tabular}[c]{@{}l@{}}9.251*\\ (5.241)\end{tabular} &
      \begin{tabular}[c]{@{}l@{}}-2.594\\ (3.676)\end{tabular} &
      \begin{tabular}[c]{@{}l@{}}1.130***\\ (0.405)\end{tabular} \\
    Xi'an &
      \begin{tabular}[c]{@{}l@{}}-4.750*\\ (2.845)\end{tabular} &
      \begin{tabular}[c]{@{}l@{}}-10.527**\\ (4.196)\end{tabular} &
      \begin{tabular}[c]{@{}l@{}}3.024\\ (2.448)\end{tabular} &
      \begin{tabular}[c]{@{}l@{}}0.883***\\ (0.242)\end{tabular} \\
    Nanjing &
      \begin{tabular}[c]{@{}l@{}}-3.511\\ (2.720)\end{tabular} &
      \begin{tabular}[c]{@{}l@{}}-6.181**\\ (2.676)\end{tabular} &
      \begin{tabular}[c]{@{}l@{}}15.003***\\ (3.368)\end{tabular} &
      \begin{tabular}[c]{@{}l@{}}0.188\\ (0.263)\end{tabular} \\
    Shanghai &
      \begin{tabular}[c]{@{}l@{}}1.406\\ (6.937)\end{tabular} &
      \begin{tabular}[c]{@{}l@{}}12.987*\\ (7.535)\end{tabular} &
      \begin{tabular}[c]{@{}l@{}}13.550**\\ (5.771)\end{tabular} &
      \begin{tabular}[c]{@{}l@{}}0.307\\ (0.601)\end{tabular} \\
    Zhengzhou &
      \begin{tabular}[c]{@{}l@{}}9.745\\ (7.425)\end{tabular} &
      \begin{tabular}[c]{@{}l@{}}-8.734\\ (9.972)\end{tabular} &
      \begin{tabular}[c]{@{}l@{}}22.730***\\ (7.751)\end{tabular} &
      \begin{tabular}[c]{@{}l@{}}1.805**\\ (0.874)\end{tabular} \\
    Wuhan &
      \begin{tabular}[c]{@{}l@{}}8.778**\\ (4.372)\end{tabular} &
      \begin{tabular}[c]{@{}l@{}}8.722\\ (5.532)\end{tabular} &
      \begin{tabular}[c]{@{}l@{}}43.851***\\ (4.057)\end{tabular} &
      \begin{tabular}[c]{@{}l@{}}-3.766***\\ (0.493)\end{tabular} \\  \hline
      \# observations &
      \begin{tabular}[c]{@{}l@{}}462\end{tabular} &
      \begin{tabular}[c]{@{}l@{}}462\end{tabular} &
      \begin{tabular}[c]{@{}l@{}}462\end{tabular} &
      \begin{tabular}[c]{@{}l@{}}462\end{tabular} \\
      \# subjects&
      \begin{tabular}[c]{@{}l@{}}10\end{tabular} &
      \begin{tabular}[c]{@{}l@{}}10\end{tabular} &
      \begin{tabular}[c]{@{}l@{}}10\end{tabular} &
      \begin{tabular}[c]{@{}l@{}}10\end{tabular} \\
      \hline
\end{tabular}}

\begin{tablenotes}
    \footnotesize
    \item[1] Standard errors are in parenthesis
    \item[2] *** p\textless{}0.01, ** p\textless{}0.05, * p\textless{}0.1
\end{tablenotes}            

\end{threeparttable}       

\label{tab:Intercept}
\end{table}

    \autoref{tab:combine_R} shows the number of new cases in one city that attracts residents' attention. It increases residents' panic (coefficient = 0.405) but has no significant influence on their confidence, while imported cases make citizens lose confidence in eradicating the disease (coefficient = -0.326). However, imported cases have no effect on people's fright. We cannot reject the null hypothesis H1-2.
    
    In terms of healthcare and city properties, it is generally accepted that spending on healthcare will improve the power to prevent coronavirus; in other words, prevention is better than cure. Our study finds that local governments' spending on healthcare prevents a panic attack and improves people's confidence in tackling the pandemic. Specifically, The higher the per capita health cost is, the higher the confidence level.

    Finally, we find people trying to escape from where the disease occurred. Conducting a robustness test whose dependent variable is population outflow to avoid a regression fallacy(see supplementary material section 2), we find a significant positive correlation between the net outflow of population and the number of local cases, which is compatible with one research under American circumstances\cite{whitaker2021did}.
    
\subsection{Robust test}
    we conduct a robustness test to avoid a regression fallacy(see appendix section 2.A). By definition, an economic model is a simplified mathematical representation of a complex interaction of economic variables, as such is built upon certain assumptions. Researchers examine how certain "core" regression coefficient estimates behave when the regression specification is modified by adding or removing regressors. The "fragility" of estimated coefficients indicates that there may be hidden errors in the model\cite{leamer1983let}. 
    
    Former study demonstrates that there is a significant positive correlation between the net outflow of population in a city and the degree of social malaise\cite{whitaker2021did}. The degree of panic caused by the epidemic is in proportion to the escape of urban dwellers (that is, the number of population outflow). It reflects the terror of urban citizens to a certain extent. According to this, our research uses the net outflow of the population as a substitute variable to supersede the variable:fear, which functions as the explained variable. Specifically, the data is obtained by subtracting the population outflow from the inflow. 
        
    The regression results show in \ref{tab:netout} which depicts the regression results of the robustness test. We find there has been a long-term mental health toll in Wuhan, the city where COVID-19 first broke out in China because people experiencing panic during an epidemic are liable to act irrationally; this phenomenon is also called a social response\cite{fast2015modelling}. Hence, we attempt to discover whether people are still escaping from the sites where disease outbreaks are occurring in the post-pandemic era. This information may help to prevent the triggering of social responses. The results of this study will also provide guidance to public health authorities to better communicate with people and provide public health responses to those who are most frightened by the virus. From the robustness test section in 5.5 and \autoref{tab:netout}, we find that people will attempt to escape from the outbreak site in the second wave of COVID-19, and that the social response between the pandemic era and the post-pandemic era is not much different.

\begin{table}[]
\caption{Regression analysis result of net outflow of the urban population}
\centering
\resizebox{\columnwidth}{!}{
\begin{tabular}{lllllll}
\hline
                     & (1)       & (2)       & (3)       & (4)       & (5)       & (6)       \\
                     & netout    & netout    & netout    & netout    & netout    & netout    \\ \hline
L.cases              & 0.013**   & 0.013**   & 0.013**   & 0.012**   & 0.012**   & 0.014**   \\
                     & (0.006)   & (0.006)   & (0.006)   & (0.006)   & (0.005)   & (0.005)   \\
L.foreign            & -0.004    & -0.004    & -0.004    & -0.005    & -0.006    &           \\
                     & (0.006)   & (0.006)   & (0.006)   & (0.006)   & (0.005)   &           \\
L.risk               & -0.003    & -0.003    & -0.003    & -0.002    &           &           \\
                     & (0.002)   & (0.002)   & (0.002)   & (0.002)   &           &           \\
L.distance           & -0.001    & -0.001    & -0.001    &           &           &           \\
                     & (0.001)   & (0.001)   & (0.001)   &           &           &           \\
pmedical             & 0.002***  & -0.003    &           &           &           &           \\
                     & (0.001)   & (0.005)   &           &           &           &           \\
density              & -0.354    &           &           &           &           &           \\
                     & (0.360)   &           &           &           &           &           \\

Chengdu              & 1.130***  & -0.424    & 0.331**   & 0.322**   & 0.312*    & 0.355**   \\
                     & (0.405)   & (1.210)   & (0.159)   & (0.163)   & (0.164)   & (0.150)   \\
Xi'an                & 0.883***  & -0.848    & 0.142     & 0.133     & 0.121     & 0.179     \\
                     & (0.242)   & (1.577)   & (0.169)   & (0.174)   & (0.175)   & (0.163)   \\
Nanjing              & 0.188     & -0.025    & 0.145     & 0.126     & 0.098     & 0.156     \\
                     & (0.263)   & (0.470)   & (0.231)   & (0.235)   & (0.235)   & (0.226)   \\
Shanghai             & 0.307     & 2.111     & 0.379***  & 0.376***  & 0.368***  & 0.395***  \\
                     & (0.601)   & (2.419)   & (0.128)   & (0.128)   & (0.129)   & (0.118)   \\
Zhengzhou            & 1.805**   & -1.315    & 0.148     & 0.124     & 0.087     & 0.138     \\
                     & (0.874)   & (2.351)   & (0.251)   & (0.258)   & (0.261)   & (0.255)   \\
Wuhan                & -3.766*** & -3.694*** & -4.146*** & -4.151*** & -4.166*** & -4.112*** \\
                     & (0.493)   & (0.559)   & (0.297)   & (0.300)   & (0.303)   & (0.291)   \\
Beijing              & -1.812*** & 2.499     & -0.307    & -0.319    & -0.331    & -0.283    \\
                     & (0.374)   & (4.098)   & (0.200)   & (0.205)   & (0.208)   & (0.189)   \\
Shenzhen             &           & 1.666     & 0.475***  & 0.462***  & 0.449***  & 0.508***  \\
                     &           & (1.693)   & (0.156)   & (0.159)   & (0.160)   & (0.133)   \\
Chongqing            &           &           & -0.169    & -0.179    & -0.191    & -0.130    \\
                     &           &           & (0.244)   & (0.250)   & (0.252)   & (0.247)   \\
date                 & -0.004    & -0.004    & -0.004    & -0.004*   & -0.004**  & -0.002    \\
                     & (0.003)   & (0.003)   & (0.003)   & (0.002)   & (0.002)   & (0.002)   \\
\_cons               & -1.586*** & 3.401     & 0.000     & 0.000     & 0.015     & -0.072    \\
                     & (0.470)   & (5.161)   & (0.250)   & (0.257)   & (0.259)   & (0.255)   \\
Obs.                 & 300       & 300       & 300       & 300       & 300       & 300       \\ \hline
\multicolumn{7}{l}{Standard errors are in parenthesis}                                       \\
\multicolumn{7}{l}{*** p\textless{}0.01, ** p\textless{}0.05, * p\textless{}0.1}            
\end{tabular}
}
\label{tab:netout}
\end{table}

\section{Discussion}

    The availability of abundant textual data from social media can be utilized to recognize people's emotional changes in the pandemic in a better way. As presented in the \autoref{tab:combine_R} and \autoref{tab:Intercept}, we can see that among these cities with Coronavirus cases, the risk perception and confidence change differently. The objective our study is to discover which kinds of information lead to risk perception or confidence. We performed a sentiment analysis based on more than 1.6 million text data. We conduct a fixed-effect regression using the analysis outcome and econometric methods.
    
    From a theoretical perspective, the findings echo and enrich cognitive theories. Prominent cognitive theories postulate that an attentional bias toward threatening information contributes to exacerbation of fear and anxiety\cite{van2014review}. Our study demonstrates people tend to seek exclusive sources of information to meet self-risk evaluation. For example, they attribute self-risk to local cases while ascribing a decline in confidence to imported cases. It might explain discrimination is produced due to the cognitive bias that exists in the information processing; therefore, some people blame foreign conspiracies for the protracted pandemic\cite{xu2021stigma,choi2020predicting}.
    
    From a methodological perspective, our research not only leverages a wealth of text data but also proposes a fusion framework combining deep multi-class sentiment analysis and regression analysis. To accurately cope with social media texts related to the pandemic, we need an efficient tool to solve the massive unstructured data. Our proposed sentiment analysis model achieved better performance than former studies. The model is also well suitable for empirical research. The model can have multiple types of outputs. It can narrow the scope of texts to a precise range. The accuracy of our model is 84.58\%, which is 7.62\% higher than that of Text-CNN\cite{zhang2015sensitivity} and 2.08\% higher than that of AC-BiLSTM\cite{liu2019bidirectional}. Next, Although combining sentiment analysis with econometric methods will contribute to reliable conclusions, a few studies have focused on model tests before an empirical analysis\cite{zhang2020does}. Our work gives a good framework for it.

    In addition, we compare the cities' differences in light of the results in \autoref{tab:Intercept}, resulting that preciously pandemic hit cities recover slowly from the suffering, which confirms the conclusion of the previous questionnaire survey: there is a long-lasting mental health toll after disasters\cite{mcfarlane1988aetiology}. This finding also confirms the riffle effect\cite{slovic1987perception}. As \autoref{tab:Intercept} shows($ coefficient = 8.778 ** t = 4.327$), residents in Wuhan, the city where the first COVID-19 breakout in China was reported, still pay great attention to it. citizens are prone to excessive anxiety and even irrational behaviors under this high-stress circumstance; this phenomenon is also called a social response\cite{fast2015modelling}. As for population outflow, citizens escape from where the disease occurred. Hence, we suggest local governments use information technology to understand where these people are going rather than rigid physical restrictions. Local governments' spending on healthcare prevents a panic attack and improves people's confidence in tackling the pandemic. One possible explanation is that governments' spending on healthcare is a sign of whether they take COVID-19 seriously and are ready for it in any case. It could explain why we observe that large cities have a better ability to respond to COVID-19. We believe that it is more worthwhile for the government to increase investment in the medical field to reduce anxiety. Public health authorities should communicate with people and provide public health responses to those who are frightened by the virus.

\section{Conclusion and limitations}
    
    In this study, we used a Python scraper to collect more than 1.6 million microblogs. Applying the ACLMM to these data, we unveil a cognitive information bias related to Coronavirus in the post-pandemic era through an econometrics method. With the coronavirus pandemic still ongoing, this work helps governments and researchers predict how public opinion will change in the next wave of the pandemic and have a better understanding of cognitive bias. Taking measures to reduce the adverse effects of this pervasive disease on people’s emotions, especially considering the mental health toll and discrimination against people entering a country in the post-pandemic era\cite{xu2021stigma}. From a methodological perspective, we propose a promising framework combining sentiment analysis using advanced deep learning technology with the empirical regression method, which can be applied in many fields such as social computation. 
    
    There are some shortcomings in the research. First, it only concerns Chinese Mainland circumstances and collects Chinese social media texts. Coronavirus statistics of Hong Kong and Taiwan is also divided into local cases and imported cases. Further, we label ambivalent texts as uncertain, leading to the sentiment analysis not being well enough. We consider integrating the aspect-level sentiment analysis method. Besides, there may be additional predictors of risk perception beyond local cases, imported cases, health costs and risk areas that should be examined. In the future, we will expand our research scope, such as considering a non-linear relationship between cases and distance results from the riffle effect\cite{slovic1987perception},\cite{ji2022effect}.

\bibliographystyle{unsrt}  
\bibliography{template}

\end{document}